\documentclass[10pt,twocolumn,letterpaper]{article}

\usepackage[pagenumbers]{cvpr} 
\usepackage{bm} 

\definecolor{cvprblue}{rgb}{0.21,0.49,0.74}
\usepackage[pagebackref,breaklinks,colorlinks,allcolors=cvprblue]{hyperref}
\newcommand{\method}{\text{SplatMesh}\xspace}

\def\paperID{9} 
\def\confName{CVPR}
\def\confYear{2025}

\title{SplatMesh: Interactive 3D Segmentation \\ and Editing Using Mesh-Based Gaussian Splatting}

\author{
Kaichen Zhou\textsuperscript{1}*
\quad Lanqing Hong\textsuperscript{3}* 
\quad Xinhai Chang\textsuperscript{1}*
\quad Yingji Zhong\textsuperscript{2} 
\quad Enze Xie\textsuperscript{3}  \\
\quad Hao Dong\textsuperscript{1}  
\quad Zhihao Li\textsuperscript{3} 
\quad Yongxin Yang\textsuperscript{4} 
\quad Zhenguo Li\textsuperscript{3} 
\quad Wei Zhang\textsuperscript{3} \\
\textsuperscript{1} Peking University \quad
\textsuperscript{2} HKUST \quad
\textsuperscript{3} Huawei Noah’s Ark Lab \quad
\textsuperscript{4} QMUL
}

\begin{document}
\maketitle
\renewcommand{\thefootnote}{\fnsymbol{footnote}}
\footnotetext[1]{Co-First Author.}
\begin{abstract}
A key challenge in fine-grained 3D-based interactive editing is the absence of an efficient representation that balances diverse modifications with high-quality view synthesis under a given memory constraint.
While 3D meshes provide robustness for various modifications, they often yield lower-quality view synthesis compared to 3D Gaussian Splatting, which, in turn, suffers from instability during extensive editing. 
A straightforward combination of these two representations results in suboptimal performance and fails to meet memory constraints.
In this paper, we introduce \method, a novel fine-grained interactive 3D segmentation and editing algorithm that integrates 3D Gaussian Splat with a precomputed mesh and could adjust the memory request based on the requirement.   Specifically, given a mesh, \method simplifies it while considering both color and shape, ensuring it meets memory constraints.
Then, \method aligns Gaussian splats with the simplified mesh by treating each triangle as a new reference point.
By segmenting and editing the simplified mesh, we can effectively edit the Gaussian splats as well, which will lead to extensive experiments on real and synthetic datasets, coupled with illustrative visual examples, highlight the superiority of our approach in terms of representation quality and editing performance.
Code of our paper can be found here: \href{https://github.com/kaichen-z/SplatMesh}{https://github.com/kaichen-z/SplatMesh}.
\end{abstract}    
\section{Introduction}
\label{sec:intro}

Editing the geometry and texture of 3D content is crucial in the computer vision community, with applications across various fields ~\cite{gao2024mesh, gao2024mani, choi2024meshgs,jambon2023nerfshop,chen2024dge,luo2024trame, yan20243dsceneeditor, salimi2025geometry, qu2025drag, xie2024sketch, gu2024dragscene, zhang2025advancing}. In Virtual Reality (VR) and Augmented Reality (AR), enhancing 3D models allows for more immersive experiences by refining object details, adjusting textures, or modifying structures in real-time~\cite{nguyen2017vremiere, tricart2017virtual}. Precise editing of 3D anatomical models aids surgical simulations, enabling doctors to visualize and adjust structures before actual procedures~\cite{cevidanes2010three, pekkan2008patient}. Artists and developers leverage 3D editing to refine character models, adjust animations, and create realistic environments in films, video games, and digital simulations~\cite{denning20153dflow}. 

\begin{figure}[t!]
    \centering
    \includegraphics[width=\linewidth, clip]{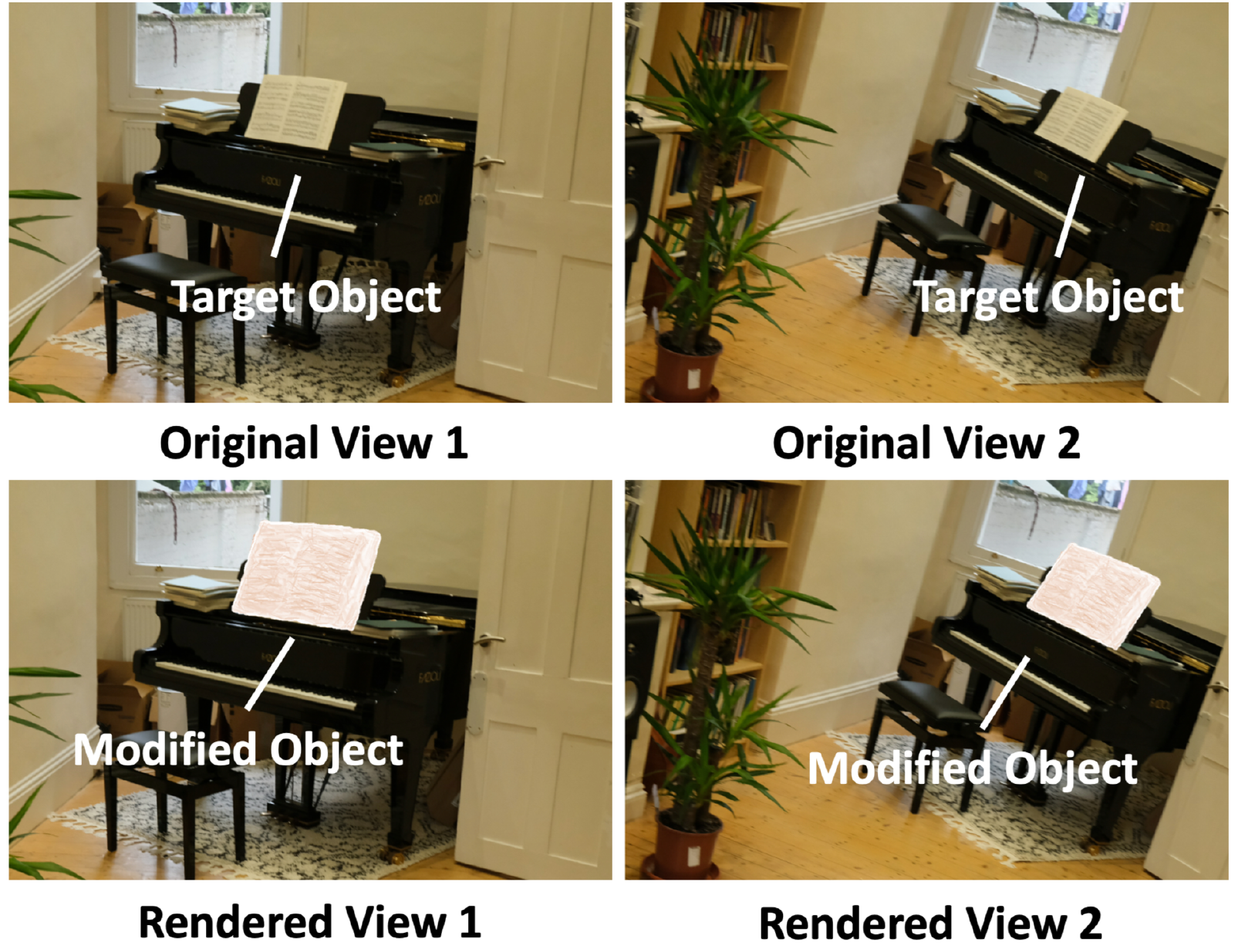}    
    \caption{\textbf{Illustration.} Our method enables high-quality interactive 3D segmentation and editing for multi-view representations. The top row displays the original images, while the row below presents the edited results, where the target book has been enlarged and stylized into a cartoon-like appearance.}
    \label{fig-brush}
    \vspace{-0.5cm}
\end{figure}

Despite the significance of 3D editing, modifying specific components within a 3D scene captured from multi-view images while ensuring consistent rendering across views and meeting memory constraints remains challenging. As illustrated in Figure ~\ref{fig-brush}, the first row depicts an indoor scene reconstructed from multi-view images, while the second row demonstrates how the multi-view representations are updated after modification to accurately reflect the change of a specific object,

Some previous multi-view editing methods rely on pretrained 2D models to perform style transfer on entire images across different views, which makes them impractical for this task~\cite{arf, chiang2022stylizing, chen2022upstnerf, huang2022stylizednerf, fan2022unified, jaganathan2024ice}.
Other previous works~\cite{gao2022get3d, sine} encode 3D information into a feature volume and perform editing by modifying the 3D feature volume. However, the low-resolution feature volume and implicit correspondences between feature volume and 3D scene, limit their applicability for precise editing, while the voxel-based representation makes them unsuitable for simulating real-world physical phenomena.

Based on these concerns, recent work has begun incorporating meshes or point clouds into the 3D editing process to achieve realistic and precise modifications. Their compact and structured representation, along with their suitability for Finite Element Methods (FEM), makes them well-suited for this purpose.
Recent work in 3D editing \cite{neumesh, yuan2022nerf} first learns a 3D mesh representation and then integrates it with a neural radiance field-like approach to achieve realistic image rendering.
However, due to the limitations of neural radiance fields, their resolution remains restricted. More advanced methods, such as those based on 3D Gaussian splitting \cite{luo20243d}, can only perform simple simulations due to the nature of splatting. By combine mesh representation with 3D gaussian splat,  \cite{gao2024mani} enables more complex modifications. However, it has two main drawbacks: it can only modify the entire image at once and requires high computational resources due to its dense sampling strategy.

In this paper, we propose a method that enables precise 3D editing while meeting specific memory requirements, which consists of representation construction stage and editing stage. \textbf{In the first stage}, given multi-view information of a 3D scenario, \method first learns a 3D mesh from the input data. \method then applies an efficient downsampling method that reduces the number of triangles in the mesh to a specific target without compromising the quality of subsequent steps. Using the downsampled mesh, \method integrates 3D Gaussian splatting to achieve realistic rendering. \textbf{In the second stage}, \method segments a specific part of the mesh from the scenario and applies target modifications, such as texture or geometry adjustments. These modifications are then transferred to the Gaussian splat using established correspondences, enabling the rendering of realistic modified images. 

Our constitution could be summarized as following:

\begin{itemize}
    \item We propose a hybrid 3D representation that integrates 3D meshes with 3D Gaussian splats. This approach achieves robust performance across a range of editing tasks while delivering high-quality view synthesis results, all within the specified memory requirements.
    \item Building on this representation, we introduce a novel 3D segmentation method that enables segmentation of both meshes and Gaussian splats using input multi-view images and a vision foundation model.
    \item Building on the proposed methods, we demonstrate two types of editing operations: geometry editing and texture editing using multi-view information. Our approach also outperforms existing methods in view synthesis and segmentation results across multiple datasets.
\end{itemize}

\section{Related Work}
\label{sec:formatting}

\paragraph{Representation for Novel View Synthesis}
Novel view synthesis generates photo-realistic images from new viewpoints using a set of posed scene captures. 
Recent developments have integrated neural networks into the rendering process, leveraging various representations such as voxels~\cite{Lombardi2019,sitzmann2019deepvoxels}, point clouds~\cite{Aliev2020,dai2020neural,zhou2023dynpoint}, multi-plane images (MPIs)~\cite{li2020crowdsampling,mildenhall2019llff,zhou2018stereo,zhou2022devnet,zhou2023manydepth2}, and implicit representations~\cite{sitzmann2019srns}. 
However, these methods often produce lower rendering quality.
A notable advancement in this area is the Neural Radiance Field (NeRF)~\cite{mildenhall2020nerf,peng2021neural,weng_humannerf_2022_cvpr,lin2021barf,yariv2021volume,wang2021neus,zhou2024neural, Yu2022MonoSDF,yu_and_fridovichkeil2021plenoxels, SunSC22,mueller2022instant, Chen2022ECCV,tang2022compressible}, which uses a Multilayer Perceptron (MLP) to encode scenes into a volumetric field. Due to the nature of latent representations, editing through NeRF is challenging.
Recently, 3DGS has gained significant attention \cite{kerbl20233d, charatan2024pixelsplat, tang2024lgm, xu2024grm, chen2024mvsplat, moon2024expressive} due to its high rendering quality.
However, compared to 3D meshes, 3DGS lack explicit 3D structure and surface definition, making them less suitable for 3D editing tasks.
Recent works \cite{guedon2024sugar, yu2024gsdf, huang20242d} integrate 3DGS into mesh structures for mesh reconstruction, enabling both view synthesis and mesh reconstruction capabilities.
While they focus primarily on mesh reconstruction, they cannot guarantee comparable quality in view synthesis.
In this paper, we propose a representation that combines the advantages of mesh and 3DGS to achieve realistic view synthesis and enable diverse editing.

\paragraph{3D Editing}
Previous editing methods typically focus on 2D editing based on single images \cite{li2020inverse, zhu2022learning, li2022physically, perez2003poisson, OM3D2014, shetty_neurips2018}. 
The advancement of 3D computer vision has made 3D editing possible, encompassing both scene-level and object-level editing. 
Scene-level editing involves modifying global attributes such as lighting \cite{guo2020object} and color palettes \cite{kuang2022palettenerf}, while intrinsic decomposition methods \cite{zhang2021nerfactor, munkberg2022extracting, hasselgren2022nvdiffrecmc, Ye2022IntrinsicNeRF, zhu2023i2, 10.1145/3588432.3591493} enable more fine-grained adjustments. 
At the object level, techniques such as Object-NeRF~\cite{yang2021objectnerf} and Liu \etal~\cite{liu2021editing} facilitate manipulation within neural radiance fields, though they are primarily restricted to rigid transformations. 
\cite{wang2024view, chen2024dge} integrate pretrained single-image editing models, such as diffusion models, with 3DGS to achieve multi-view editing. However, their performance is constrained by the limitations of the pretrained models and lacks precise editing capabilities.
NeuMesh~\cite{neumesh} represents a significant advancement in object-level editing with its fine-grained control; however, its mesh-based representation constrains its performance in view synthesis.
Despite these advancements, a major drawback in the field remains the optimization for efficiency, with many methods requiring extensive optimization and inference times for practical editing applications.
Concurrent works, namely GaMeS \cite{gao2024mesh}, Mesh-GS \cite{gao2024mani} and Meshgs \cite{choi2024meshgs}, rely on a relatively large number of splats since their sampling process is performed on a per-triangle basis via pre-computed mesh. Additionally, these methods only associate the position of 3D Gaussian splats with the mesh without offering an efficient interactive segmentation mechanism. In contrast, \method is designed to deliver an interactive system capable of supporting diverse and fine-grained editing tasks.
\section{Methodology}
\begin{figure*}
    \centering
    \includegraphics[width=0.98\textwidth]{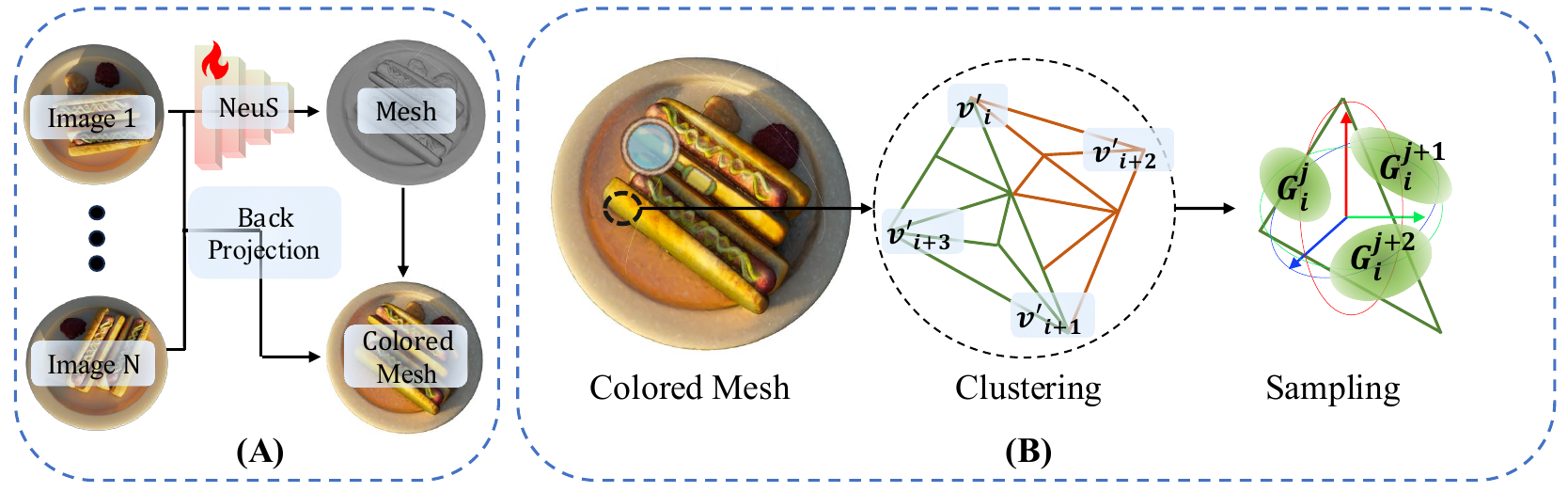}
    \caption{\textbf{Illustration of Mesh Construction and Surface Rendering in \method.}
    The \method involves 2 steps. In Step (1), a multi-view approach is employed to reconstruct the mesh, followed by projecting the color onto the mesh to create the colored mesh representation. In Step (2), the colored mesh is initially downsampled based on geometry and texture information, and 3DGS is then sampled within the new set of vertices.}
    \label{fig:mesh-construction}
\end{figure*}

The user is provided with a set of multi-view images $\{ \bm{I}_1, \bm{I}_2, ..., \bm{I}_N \}$ captured from different viewpoints $\{ \bm{C}_1, \bm{C}_2, ..., \bm{C}_N \}$, depicting a 3D scene. The editing process begins with the user selecting a random image $\bm{I}_{u}$ and segmenting the region of interest using prompts. The user then specifies editing instructions $\bm{I}_{u}^e$ for the segmented region, enabling targeted modifications.

The goal of \method is to generate a set of modified multi-view images $\{ \bm{I}_1^e, \bm{I}_2^e, ..., \bm{I}_N^e \}$, ensuring that the modifications applied to $\bm{I}_{u}^e$ are reflected consistently across all views, resulting in a coherent and realistic set of modified images. This methodology comprises four key steps:

1. **Integrating Mesh and Gaussian Splats**:  
   We utilize a state-of-the-art multi-view geometry estimation algorithm to reconstruct a mesh representation $\bm{M}$ of the 3D scene. Based on the geometric and texture features of $\bm{M}$, we downsample the original set of vertices $\{\bm{v}_i\}$ to new set $\{\bm{v'}_i\}$. For each new triangle $\{\bm{v'}_i, \bm{v'}_{i+1}, \bm{v'}_{i+2}\}$, we sample $N^s$ 3DGS. These 3DGS form the basis for subsequent view synthesis and editing processes.

2. **3D Mesh Segmentation**:  
   To facilitate accurate editing, we construct a 3D mask $\mathcal{M}$ that corresponds to the edited segmented image $\bm{I}_{u}^e$. This segmentation process ensures alignment between the user's 2D editing inputs and the underlying 3D representation.

3. **View Synthesis and Editing**:  
   Leveraging the proposed view synthesis model and 3D segmentation framework, we demonstrate two editing modalities: geometric editing and texture editing. These operations ensure the edits applied to $\bm{I}_{u}^e$ are propagated consistently across all viewpoints, maintaining coherence and realism.

\subsection{Preliminary}

In 3D Gaussian Splatting (3DGS), rendering is performed using explicit 3DGS as the main primitives. A 3D Gaussian point, represented mathematically, is defined as:

\begin{equation}
G(\boldsymbol{x}) = \exp\left( -\frac{1}{2}(\boldsymbol{x}-\boldsymbol{\mu})^\top \Sigma^{-1} (\boldsymbol{x}-\boldsymbol{\mu}) \right),
\label{eq:3d_gaussian}
\end{equation}

where $\boldsymbol{x}$ denotes a point in 3D space, $\boldsymbol{\mu}$ is the mean position, and $\Sigma$ is the covariance matrix that governs the spread of the Gaussian. Each Gaussian also has a view-dependent color $\boldsymbol{c}$ and an opacity value $\boldsymbol{o}$, with the color typically represented using spherical harmonics (SH). $\Sigma$ is parameterized by a unit quaternion $\boldsymbol{q}$ and a 3D scaling vector $\boldsymbol{s}$, where:$\Sigma = \boldsymbol{R}\boldsymbol{S} \boldsymbol{S}^\top \boldsymbol{R}^\top.$ When rendering an image from a specific viewpoint, the 3D Gaussians are projected onto the image plane, transforming them into 2D Gaussians. The resulting 2D covariance matrix is computed as:
$\Sigma' = \boldsymbol{J}\boldsymbol{W}\Sigma \boldsymbol{W}^\top \boldsymbol{J}^\top,$
where $\boldsymbol{W}$ is the viewing transformation matrix, and $\boldsymbol{J}$ is the Jacobian of the affine approximation of the perspective projection. The means of the 2D Gaussians are computed via the projection matrix, while the pixel colors are obtained through alpha blending.
For each pixel in the image, the color $\mathcal{C}$ is derived by combining the contributions of the $N$ ordered 2D Gaussians as follows:

\begin{equation}
\mathcal{C} = \sum_{i \in \{N\}} T_{i} \alpha_{i} \boldsymbol{c}_{i} \quad \text{where} \quad T_i = \prod_{j=1}^{i-1} (1 - \alpha_j),
\label{eq:alpha_blending}
\end{equation}

Here, the opacity $\alpha_{i}$ for each Gaussian is determined by multiplying the opacity $\boldsymbol{o}$ with the probability based on the 2D covariance $\Sigma'$ and the pixel’s position in the image.

\subsection{Mesh-based View Synthesis}
\label{sec:binding}

In this section, we propose integrating 3DGS with a precomputed mesh. By manipulating the mesh, we enable modifications to the 3DGS, which in turn affect the synthesized frames. Unlike Sugar\cite{guedon2024sugar}, which uses the mesh solely to initialize 3D splats, our approach establishes both color and geometric relationships between the mesh and the 3DGS. This integration involves two steps, as illustrated in Figure~\ref{fig:mesh-construction}. In step (A), we utilize the reconstructed mesh and image information to generate a colored mesh.
In step (B), we downsample the original set of mesh vertices into a smaller set and then sample 3DGS based on new set.

\subsubsection{Mesh Coloring}

To achieve mesh coloring, we follow the following steps: (1) Given a frame $\bm{I}_n$ and the camera matrix $\bm{C}_n$, we estimate the corresponding depth $\bm{D}'_n$ based on the mesh $\bm{M}$. This depth information is used to compute the corresponding 3D point $\bm{P}$ for each pixel $\bm{p}$ using the equation $\bm{P} = \bm{D}'_n(\bm{p}) \bm{K}_n \bm{C}_n \bm{p}$, where $\bm{K}_n$ represents the intrinsic camera matrix. (2) By reprojecting all the pixel-wise color information into the 3D point space, we obtain a set of colored points $\{\bm{P}_n\}_{n=1}^{N}$, where $\bm{P}_n$ represents the 3D points associated with frame $\bm{I}_n$. (3) For each vertex $\bm{v}$ in the mesh $\bm{M}$, we find the $K$ closest points from the colored points obtained in the previous step. These closest points serve as the neighboring points for vertex $\bm{v}$: \begin{equation}
    \{\bm{P}_k\}_{k=1}^{k=K} = argmin_K (\bm{v} - \bm{P})^2.
\end{equation}
By performing these steps, we can color feature vectors to each vertex in the mesh, enabling the construction of the colored mesh representation $\overline{\bm{M}}$. The feature of vertice $\bm{v}$ could be written as:
\begin{equation}
    \bm{c}(\bm{v}) = \sum_{k=1}^{k=K} w_k \mathcal{C}(\bm{P}_k) / \sum_{k=1}^{k=K} w_k, \;\; w_k=\frac{1}{||\bm{P}_k-\bm{v}||},
\label{eqn:weight_sum}
\end{equation}
where $w_k$ is the weight based on the inversed distance.

\subsubsection{Geometry and Texture-Guided 3D Point Sampling}

Given the colored mesh representation $\overline{\bm{M}}$. Instead of uniformly sampling $N^s$ splats for each triangle, \method firstly downsample vertices into new set of vertices $\{ \bm{v'}_i \}_{i=1}^{N'}$. Then \method generate $N^s$ sample for each new triangle. 

\textbf{Downsamplig:} To achieve efficient mesh downsampling and obtain the downsampled vertices $\{ \bm{v'}_i \}_{i=1}^{N'}$ from the original vertex set $\{ \bm{v}_i\}_{i=1}^{N}$, we employ Quadric Error Metric (QEM), which enables accurate mesh simplification.  

Traditional QEM is formulated as:  
\begin{equation}
    E(\bm{v'}) = \bm{v'}^T Q' \bm{v'}
\end{equation}
where $Q' = Q_1 + Q_2$, and $Q_1$ and $Q_2$ represent the distance metrics of vertices $\bm{v}_1$ and $\bm{v}_2$ to the corresponding triangle planes. The error quadric $Q$ for a given plane $ax + by + cz + d = 0$ is defined as:  
\begin{equation}
Q = \bm{n} \bm{n}^T, \quad \text{where} \quad \bm{n} = [a, b, c, d]^T.
\end{equation}

However, this method only considers geometric features and does not account for additional properties such as color and smoothness.  
Inspired by previous work~\cite{745312}, we extend the color dimensions into the original coordinate space, redefining each vertex as \(\bm{\overline{v}} = (x, y, z, r, g, b)\), and construct the corresponding quadric matrix \(\overline{Q}\) accordingly. Besides, noting that QEM struggles to maintain high triangle quality and may weaken less prominent features~\cite{10.1145/3658159}, we introduce a post-processing step using Taubin smoothing~\cite{10.1145/218380.218473}. This additional procedure ensures a more uniform distribution of points across the mesh, enhancing both geometric and visual fidelity in the downsampled mesh.

\textbf{Sampling:} In this step, we aim to associate new triangle $\{\bm{v'}_i, \bm{v'}_{i+1}, \bm{v'}_{i+2}\}$ with $N^s$ gaussian splats $\bm{G}_{i}^j$. The core idea is to leverage the local coordinate system of each triangle, which is defined by three fundamental directions:
\begin{equation}
    \begin{aligned}
        \bm{R}_i = \Bigg[ &\bm{d}_2 = \frac{(\bm{v'}_{i+1} - \bm{v'}_{i}) \times  (\bm{v'}_{i+2} - \bm{v'}_{i})}{ \lVert (\bm{v'}_{i+1} - \bm{v'}_{i}) \times  (\bm{v'}_{i+2} - \bm{v'}_{i}) \rVert}, \\
        & \bm{d}_1 = \frac{(\bm{v'}_{i+1} - \bm{v'}_{i})}{ \lVert (\bm{v'}_{i+1} - \bm{v'}_{i}) \rVert}, \bm{d}_3 = \bm{d}_1 \times \bm{d}_2 \Bigg].
    \end{aligned}
\end{equation}

The center of this coordinate system is given by
\begin{equation}
    \bm{\mu_i} = (1 - \sqrt{r_1})\bm{v'}_{i} + \sqrt{r_1}(1 - r_2)\bm{v'}_{i+1} + \sqrt{r_1}r_2\bm{v'}_{i+2},
\end{equation}
where $r_1 \& r_2$ are randomly sampled from the interval $[0,1]$.

The color of this coordinate is expressed as
\begin{equation}
    \bm{c}_i = (1 - \sqrt{r_1})\bm{c}(\bm{v'}_{i}) + \sqrt{r_1}(1 - r_2)\bm{c}(\bm{v'}_{i+1}) + \sqrt{r_1}r_2\bm{c}(\bm{v'}_{i+2}).
\end{equation}

The scale of this coordinate is represented by
\begin{equation}
    \bm{s}_i = [\lVert \bm{v'}_{i+1} - \bm{v'}_{i} \rVert, \lVert \bm{v'}_{i+2} - \bm{v'}_{i} \rVert, \lVert  \bm{v'}_{i+2} - \bm{v'}_{i+1} \rVert].
\end{equation}

During the optimization process, the parameters of a Gaussian splat are updated as follows:
\begin{equation}
    \begin{aligned}
     & \tilde{\bm{R}_i} = \hat{\bm{R}_i} \bm{R}_i , \quad \tilde{\bm{\mu}_i} = \bm{\mu}_i + \bm{R}_i \hat{\bm{\mu}_i}, \\  &\quad \tilde{\bm{s}_i} = \bm{s}_i \hat{\bm{s}_i}, \quad \tilde{\bm{c}_i} = \bm{c}_i + \hat{\bm{c}_i}.
    \end{aligned}
\end{equation}
where $\hat{\bm{R}_i}$, $\hat{\bm{\mu}_i}$, $\hat{\bm{s}_i}$, and $\hat{\bm{c}_i}$ are the optimizable parameters; $\tilde{\bm{R}_i}$, $\tilde{\bm{\mu}_i}$, $\tilde{\bm{s}_i}$, and $\tilde{\bm{c}_i}$ are the parameters for Gaussian splats.

\begin{figure*}
    \centering
    \includegraphics[width=1.\textwidth]{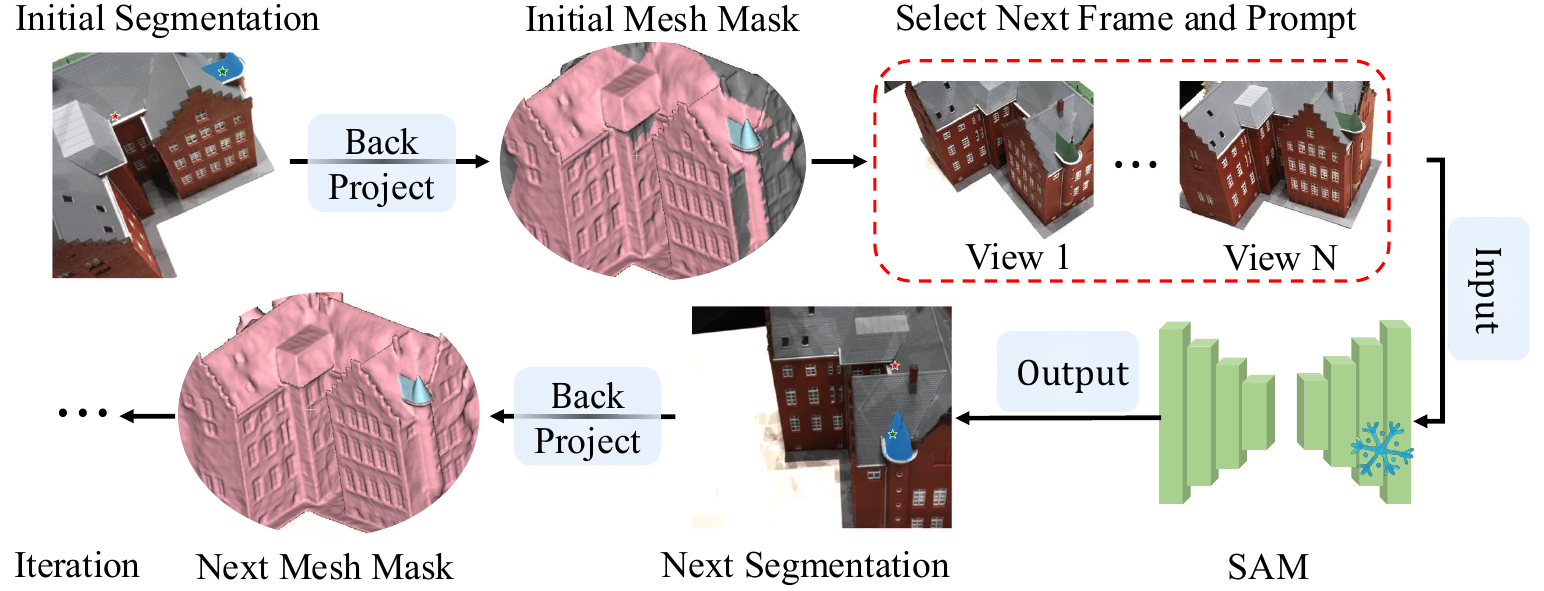}
    \caption{\textbf{Illustration of 3D Segmentation Pipeline}. The red and green stars represent $\bm{pr}^{oth}_{new}$ and  $\bm{pr}^{obj}_{new}$, used for the SAM model.}
    \label{fig:mesh-segmentation}
\end{figure*}

\subsection{3D Mesh Segmentation}
\label{sec:segment}

To enable precise 3D editing using the segmented frame $\bm{I}_u^e$, we extend the 2D mask to 3D. Accurate 3D segmentation helps modify specific parts while keeping others static. Unlike 2D methods like SAM, interactive 3D segmentation is limited. Though SAM3D \cite{cen2023segment} exists, it struggles with fine-tuning for prompt updates. We propose an efficient 3D segmentation approach based on 2D supervision, as illustrated in Figure~\ref{fig:mesh-segmentation}.

\begin{itemize}
    \item \textbf{(a) 2D Mask Generation:} The 2D mask $\bm{m}_{pre} = \{\bm{p}_j\}, \; \bm{p}_j \in \{0,1\}$ for image $\bm{I}_{pre}$ is generated using user-provided prompts $\bm{pr}^{obj}_{pre}$ and $\bm{pr}^{oth}_{pre}$ with a trained SAM.
    
    \item \textbf{(b) 3D Masking:} The 2D mask $\bm{m}_{pre}$ is back-projected to obtain 3D masked vertices $\mathcal{M} = \{\bm{v}_j\}$. 
    
    \item \textbf{(c) Next Frame Selection:} The next frame is chosen as:
    \begin{equation}
        \bm{I}_{C} = \underset{\bm{C}}{\arg\max} \; \langle \mathcal{V}_{\bm{C}}, \mathcal{M} \rangle.
    \end{equation}
    where $\langle x, y \rangle$ denotes the overlap between $x$ and $y$.
    
    \item \textbf{(d) New Prompt Generation:} Given accurate prior prompts, new prompts should remain close to $\bm{pr}_{pre}$. We determine $\bm{v}^{obj}_{new}$ as:
    \begin{equation}
        \bm{v}^{obj}_{new} = \arg\min \; (d),
    \end{equation}
    \begin{equation}
        d = \sum_{pre}||\mathcal{V}_{\bm{C}} [Observed \; Part] - \bm{v}^{obj}_{pre}||.
    \end{equation}
    Similarly, we find $\bm{v}^{oth}_{new}$, project them to 2D, and generate pixel-wise prompts $\bm{pr}^{oth}_{new}$ and $\bm{pr}^{obj}_{new}$. The new mask $\bm{m}_{new}$ is obtained via SAM, updating $\bm{m}_{pre}=\bm{m}_{new}$, and restarting from step (b).
\end{itemize}

For robustness, in step (b), direct replacement of 3D vertices using $\bm{m}_{pre}$ may lead to segmentation errors. Instead, for each vertex $\bm{v}$, segmentation results across prior frames $\{\mathcal{M}_n\}_{n=1}^N$ are considered. If most frames classify $\bm{v}$ as $1$, it is marked as an 'object'; otherwise, as 'others'. This ensures consistent segmentation across frames. The final 3D mask $\mathcal{M}$ is then projected onto each frame for precise segmentation, enabling further analysis and editing.

\subsection{3D Editing}
\label{sec:editing}
Building upon the previously proposed Neural Mesh Construction and 3D Mesh Segmentation methods, we will now focus on implementing two specific types of editing. It's important to note that although we only present these two editing methods, our approach can be extended to other editing operations, such as scaling.


\paragraph{Geometry Deformation.}
Our representation is primarily based on an explicit mesh representation, which enables us to achieve consistent multi-view geometric editing by manipulating the 3D neural mesh. In order to maintain local consistency within the 3D representation, we have deviated from the traditional volume rendering process, which relies on global positions. Instead, we calculate local normals and relevant distances between queried points and vertices. This allows us to infer the structure and appearance information of the 3D scenario. Our Experiment section includes several examples that demonstrate these concepts.

In our proposed single-view-based 3D editing approach, we first utilize a 3D segmentation method to create a mask for vertices that are subject to deformation and vertices that are intended to remain static. Next, we employ a mesh deformation algorithm, such as ARAP (As-Rigid-As-Possible), to transform the shape of the 3D mesh. Finally, the 3D neural mesh is used to construct the 3D appearance information. 

\paragraph{Texture Painting.}
In addition to 3D geometry deformation, we introduce texture painting operations for 3D editing. This allows users to draw directly on a selected frame $\bm{I}_r^e$ from the provided frames ${\bm{I}_1, \bm{I}_2, ..., \bm{I}_N}$. Instead of modifying the UV mapping, our approach, called \method, offers a more intuitive way to perform texture painting.

To begin the texture painting process, we start with the edited frame $\bm{I}_u^e$. First, we segment the edited region $\bm{m}_e$ by comparing $\bm{I}_u^e$ with the corresponding original frame $\bm{I}_u$. This segmentation helps identify the specific area that has been modified.
We then back-project the 2D mask $\bm{m}_e$ into 3D vertices, resulting in the vertex mask $\mathcal{M}$. This vertex mask indicates which vertices of the colored mesh $\overline{\bm{M}}$ are affected by the edited region.
Using the vertex mask $\mathcal{M}$, we modify the colored information of the mesh $\overline{\bm{M}}$ for the masked-out vertices. This ensures that the texture changes applied to the edited region are reflected in the 3D representation. Finally, based on the edited mesh $\overline{\bm{M}}^e$, we fine-tune our 3D gaussian. 
By incorporating texture painting operations into our approach, we provide users with the ability to directly manipulate and enhance the visual details of 3D objects. This expands the range of editing possibilities and offers a more intuitive and natural way to modify textures without the need for complex UV mapping adjustments.

\section{Experiments}


\begin{figure*}[t!]
    \centering
    \vspace{-1em}
    \includegraphics[width=\linewidth, clip]{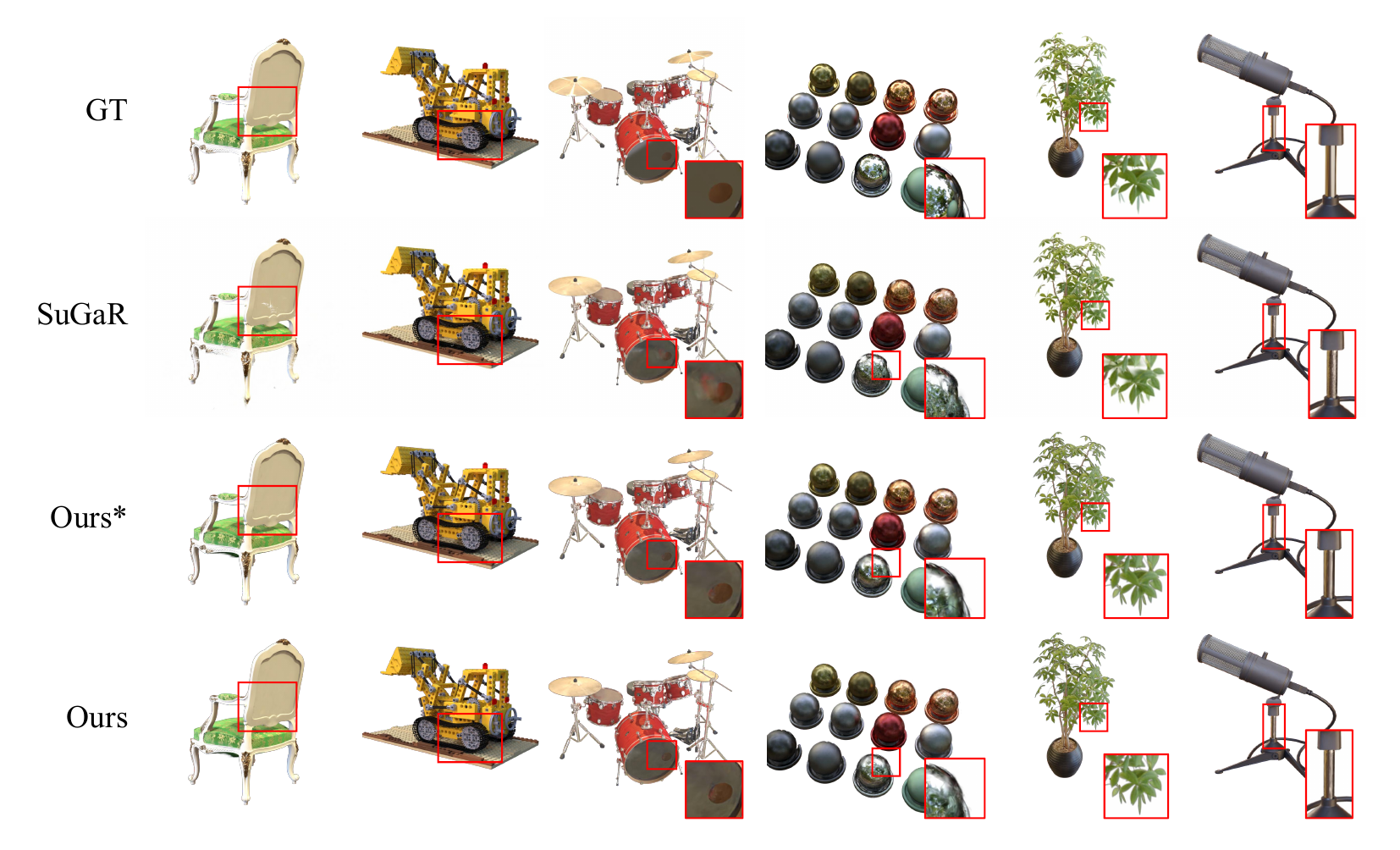}
    \vspace{-0.6cm}
    \caption{\textbf{Qualitative Comparison of View Synthesis on NeRF Synthetic Dataset.}
    We conducted a comparative analysis with SuGaR, evaluating the outcomes of extracting 3DGS from meshes and rendering test perspectives under identical fine-tuning iterations. Our* denotes the configuration employing our mesh downsampling method with a 1/3 face retention ratio and Our denotes the configuration w/o using downsampling. In contrast to SuGaR, our model demonstrates a significant reduction in artifact generation and yields results with enhanced granularity.
    }
    \vspace{-0.3cm}
    \label{fig:nerf_results}
\end{figure*}

\begin{table*}[tb]
\centering
\resizebox{1.0\linewidth}{!}{
\tabcolsep 3pt
  \begin{tabular}{@{}lcccccccccc@{}}
    \toprule
    \multicolumn{1}{c}{} & \multicolumn{3}{c}{Sugar (811K)} & \multicolumn{3}{c}{Ours* (94K)} & \multicolumn{3}{c}{Ours (284K)} \\
    \cmidrule(r){2-4} \cmidrule(r){5-7} \cmidrule(r){8-10}
    & PSNR $\uparrow$ & SSIM $\uparrow$ & LPIPS $\downarrow$ & PSNR $\uparrow$ & SSIM $\uparrow$ & LPIPS $\downarrow$ & PSNR $\uparrow$ & SSIM $\uparrow$ & LPIPS $\downarrow$ \\
    \midrule
    chair & 33.16 & 0.979 & 0.023 & 33.83 & 0.980 & 0.020 & \textbf{34.87} & \textbf{0.985} & \textbf{0.013} \\
    drums & 25.28 & 0.946 & 0.048 & 25.93 & 0.953 & 0.041 & \textbf{25.97} & \textbf{0.954} & \textbf{0.039} \\
    ficus & 32.39 & 0.981 & 0.019 & \textbf{35.18} & \textbf{0.986} & \textbf{0.013} & 35.11 & \textbf{0.986} & \textbf{0.013} \\
    hotdog & 36.22 & 0.983 & 0.023 & 36.08 & 0.980 & 0.030 & \textbf{36.77} & \textbf{0.984} & \textbf{0.021} \\
    lego & 33.67 & 0.976 & 0.023 & 34.72 & 0.978 & 0.024 & \textbf{36.01} & \textbf{0.983} & \textbf{0.015} \\
    materials & 27.32 & 0.940 & 0.048 & 28.44 & 0.948 & 0.055 & \textbf{28.81} & \textbf{0.953} & \textbf{0.045} \\
    mic & 34.61 & 0.991 & 0.008 & 36.23 & 0.992 & 0.007 & \textbf{36.51} & \textbf{0.992} & \textbf{0.007} \\
    ship & 29.18 & 0.883 & 0.111 & 30.45 & 0.896 & 0.119 & \textbf{30.86} & \textbf{0.898} & \textbf{0.106} \\
    \midrule
    Average & 31.48 & 0.960 & 0.038 & 32.61 & 0.964 & 0.039 & \textbf{33.11} & \textbf{0.967} & \textbf{0.032} \\
    \bottomrule
  \end{tabular}
}
\caption{\textbf{Quantitative Results for View Synthesis on NeRF Synthetic.} Performance evaluation of Sugar and \method for novel view synthesis, with all methods trained for 10,000 iterations. Our* denotes the configuration employing our mesh downsampling method with a 1/3 face retention ratio, and best results are emphasized in boldface.}
\vspace{-0.5cm}
\label{tab:mesh_nerf}
\end{table*}

To demonstrate the advantages of our method, we conduct a series of experiments aimed at providing comprehensive evidence of its capabilities in various aspects, including (1) novel view synthesis, (2) 3D mesh segmentation, and (3) 3D interactive editing.
Detailed experimental settings and additional visualization results are provided in the Appendix.

\paragraph{Datasets.} Our experiments are conducted on various datasets, including DTU \cite{jensen2014large} and NeRF Synthetic \cite{NeRF}. 
For the DTU dataset \cite{jensen2014large}, we used the IDR \cite{yariv2020multiview} configuration, utilizing 15 scenes with images of 1600 × 1200 resolution and accompanying foreground masks. 
To facilitate metric evaluation for both rendering and mesh quality, we randomly selected $10\%$ of the images as a test split and used the remaining images for training. 
For the NeRF Synthetic dataset, we adhered to the official split guidelines. 
Due to page limitations, the results of our method on the DTU dataset are provided in the appendix.

\subsection{View Synthesis}
\label{exp1}

We first conduct an evaluation of view synthesis performance across several surface rendering techniques. 
We include comparisons with Sugar~\cite{guedon2024sugar}. 
The experimental setup follows~\citet{xiang2021neutex}, assessing each method on two benchmark datasets, i.e., DTU and NeRF 360$^{\circ}$ Synthetic with three metrics, namely PSNR, SSIM, and LPIPS.

\textbf{Results.} As shown in Table~\ref{tab:mesh_nerf}, a comparison between Sugar and Ours* reveals that our method achieves superior view synthesis results with significantly fewer points. Specifically, using only $35\%$ of the points, our approach surpasses Sugar by $5\%$ in terms of PSNR. Additionally, Our* employs a geometry-texture fused mesh downsampling method, reducing the point count to $12\%$ of Sugar's, while still achieving superior performance. This demonstrates that our method provides an efficient initialization for Gaussian points, as increasing the number of points does not lead to noticeable performance gains. The qualitative results in Figure~\ref{fig:nerf_results} further corroborate these findings, showcasing that our method effectively preserves detailed geometry and texture information in the generated images.

We also evaluated our approach with the DTU dataset, as presented in Table 1 of the Appendix and Figure 1 of the Appendix. On this dataset, Ours achieves a $10\%$ improvement in PSNR, a $10\%$ improvement in SSIM, and a $60\%$ reduction in LPIPS compared to Sugar, while utilizing only $9\%$ of the points. Our* attains comparable quality with only $1/3$ of the points required by baseline approaches. Qualitatively, our results exhibit minimal blurring or unclear regions, further emphasizing the effectiveness of our method.

\begin{figure}[t!]
    \centering
    \vspace{-1em}
    \includegraphics[width=\linewidth, clip]{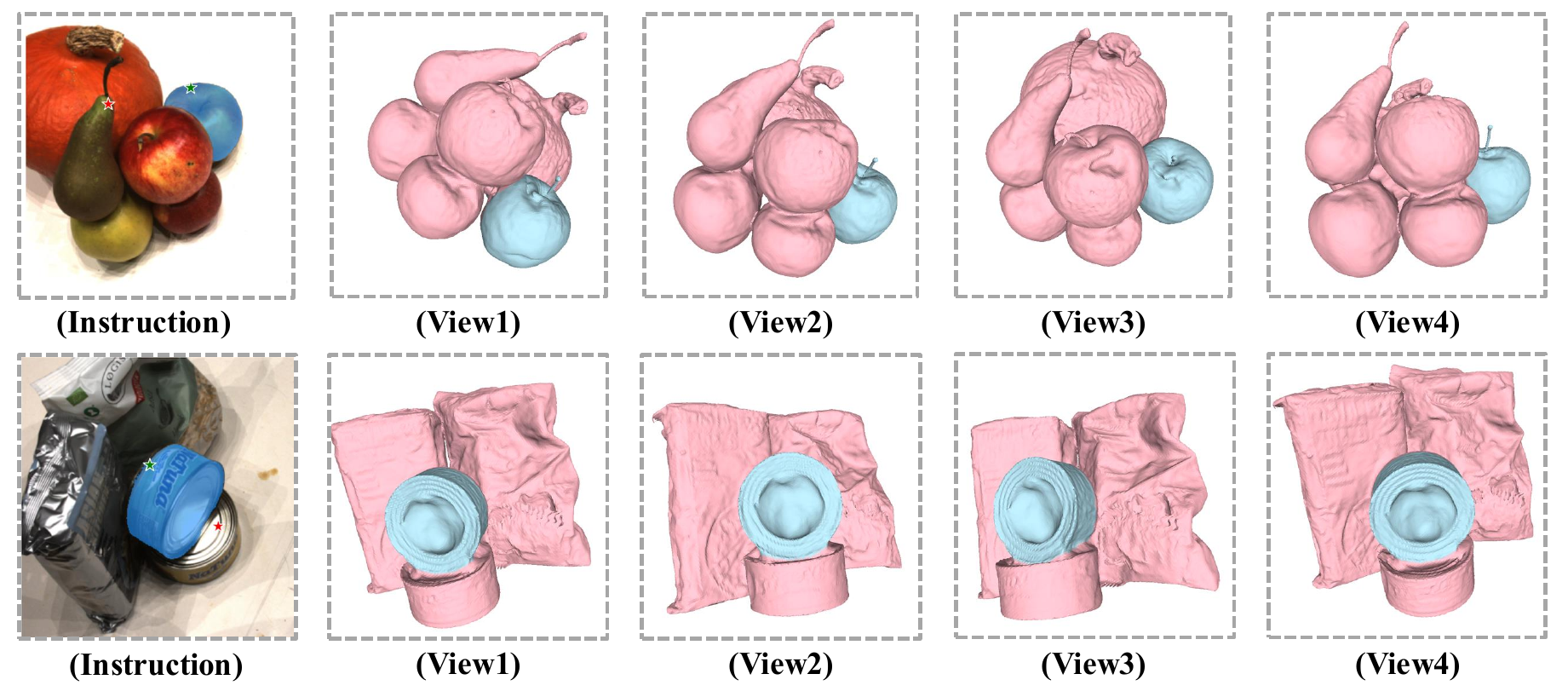}
    \vspace{-0.3cm}
    \caption{\textbf{Example of 3D Segmentation with \method.}
    Our approach enables interactive 3D segmentation using 3D prompts. The first column displays the positions of 2D prompts, while the remaining columns present the 3D segmentation results.
    }
    \label{fig:interative_segment}
\end{figure}

\subsection{3D Interactive Segmentation}
\label{exp2}

In this section, we present the results of our study on 3D mesh segmentation, leveraging interactive user-provided instructions to guide the process. The primary goal is to evaluate the effectiveness of our proposed method in accurately segmenting reconstructed 3D meshes based on 2D input prompts. By integrating user input into the segmentation pipeline, we aim to showcase the ability of our method to bridge 2D annotations and 3D geometry, enabling precise and context-aware segmentation.

As illustrated in Figure~\ref{fig:interative_segment}, our method demonstrates a robust ability to propagate segmentation prompts provided in a single 2D image to the corresponding 3D mesh. By mapping user-defined masks from the 2D domain into 3D, we achieve accurate and comprehensive segmentation of the target object within the reconstructed mesh. This approach ensures that the 3D mask faithfully reflects the user’s intent, making it highly suitable for tasks requiring precision.

\begin{figure}[t!]
    \centering
    \vspace{-0.5em}
    \includegraphics[width=\linewidth, clip]{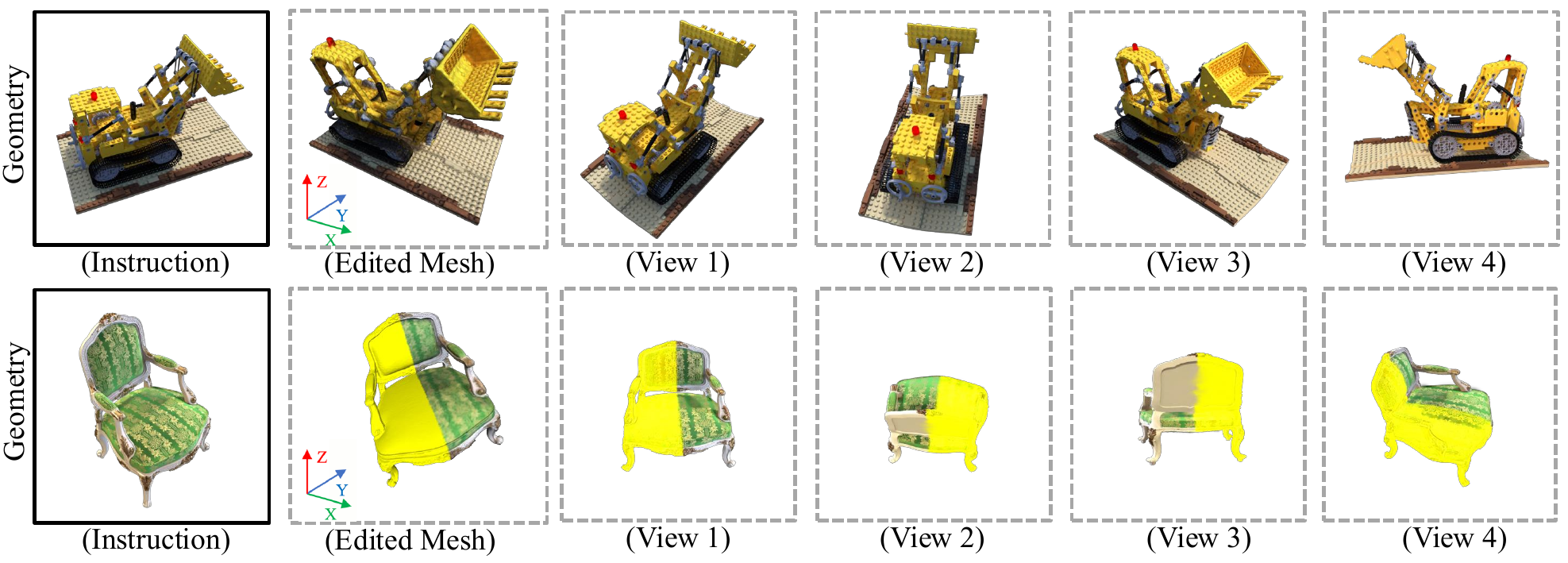}
    \caption{\textbf{Example of 3D Interactive Editing with \method.}
    Our approach allows for interactive geometry editing (1st row) and interactive appearance editing without fine-tuning (2nd row).
    }
    \label{fig:interative editing}
\end{figure}

\subsection{3D Interactive Editing}

\paragraph{Interactive Geometry Editing.}
In this section, we demonstrate the effectiveness of our method in interactive 3D geometry editing. As illustrated in Figure~\ref{fig:interative editing} (specifically in the first row), our approach enables users to manipulate an object within a view interactively. This manipulation directly alters the corresponding 3D mesh, leading to coherent and synchronized changes across different viewpoints. This experiment highlights our algorithm's capability not only in generating realistic and visually appealing views but also in maintaining the consistency of the reconstructed mesh during dynamic geometry modifications. Such interactive editing underscores the practical utility of our method in applications requiring real-time 3D scene manipulation and visualization.

\paragraph{Interactive Appearance Editing.}
We further demonstrate the capabilities of our method in interactive surface editing. 
Our approach enables users to modify an object's surface within a view. 
For example, in one of our experiments, we modify the object's surface by coloring half of the mesh yellow, as shown in the second row of Figure~\ref{fig:interative editing}.
This modification is reflected in the corresponding mesh, leading to changes in the reconstructed views from new perspectives. 
The experiment showcases how our algorithm facilitates not only the modification of the mesh's appearance but also ensures that these changes are consistently translated across different views. The result is visually compelling and customized outputs, illustrating the practical utility of our method in applications that require detailed and personalized surface editing.

\begin{table}[tb]
  \label{tab:nvos}
  \scriptsize
  \centering
  \begin{tabular}{@{}lccccc@{}}
    \toprule
    & Base & QEM (1/3) & QEM+ (1/3) & QEM+ (1/6) & QEM+ (1/9) \\
    \midrule
    Average & 33.11 & 32.53 & 32.61 & 31.92 & 31.39 \\
    \bottomrule
  \end{tabular}
  \caption{\textbf{Quantitative Comparison of Geometry Compression Strategies.} Average PSNR results on NeRF Synthetic demonstrating the effectiveness of our quantization-enhanced methods (QEM variants) versus the baseline approach. The suffix ratios denote parameter reduction factors.}
  \label{table.ablation}
  \vspace{-0.1cm}
\end{table}

\begin{figure}[t!]
    \centering
    \includegraphics[width=\linewidth, clip]{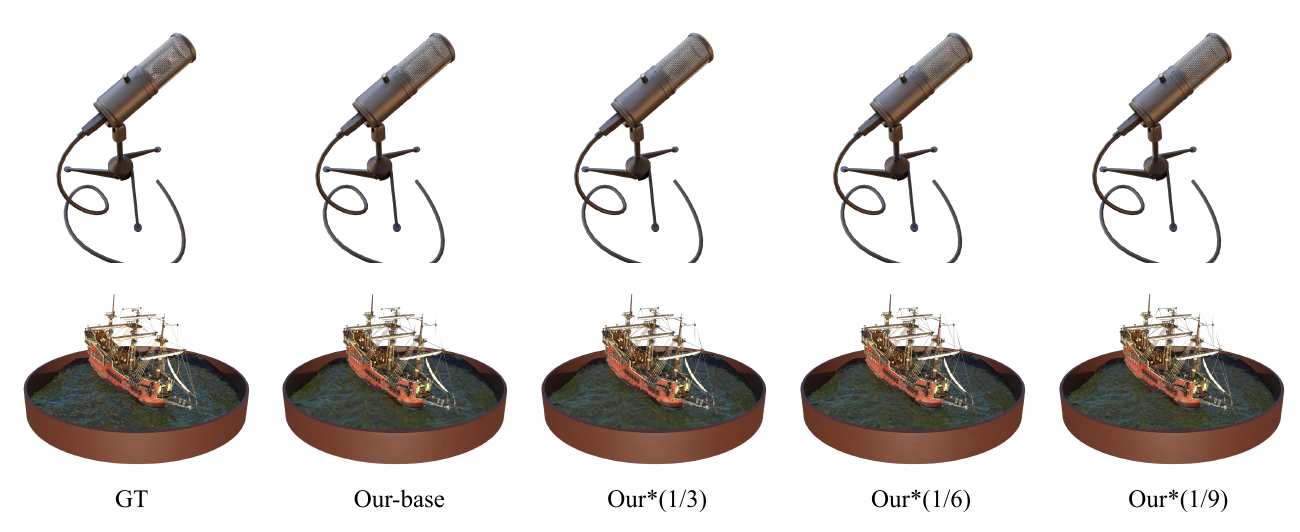} 
    \caption{\textbf{Qualitative Results of the Ablation Study.} Comparative evaluation of mesh downsampling with face retention ratios (1/3, 1/6, 1/9) demonstrates effective preservation of visual fidelity for both isolated objects and complex scene geometries under high compression ratios.}
    \vspace{-0.3cm}
    \label{fig:ablation_study}
\end{figure}
\subsection{Ablation Studies}
\label{exp5}
To further validate the effectiveness of \method, we conduct ablation studies focusing on two critical aspects: (1) the impact of integrating geometry-texture information with smoothing (QEM+) , compared to geometry-only methods (QEM), and (2) variations in the mesh downsampling ratio. Quantitative results are presented in Table~\ref{table.ablation}, and qualitative outcomes are illustrated in Figure~\ref{fig:ablation_study}.

Several important observations emerge from this study. First, incorporating texture information and smoothness-aware optimization (QEM+) improves mesh downsampling guidance compared to geometry-only QEM, enabling better Gaussian initialization. Besides, thanks to its effective initialization, \method maintains comparable rendering quality for both single objects and complex scenes even after aggressive point reduction. Even when reducing points to $1/9$ of the base configuration, our QEM+ strategy maintains rendering quality with $<2\%$ PSNR degradation, demonstrating remarkable efficiency. 

\section{Conclusion}

In this study, we present \method, a highly efficient algorithm for interactive 3D segmentation and editing, designed to operate seamlessly without fine-tuning for user-specific prompts. By integrating mesh representation with 3D Gaussian splitting, \method enables accurate 3D segmentation and interactive editing with remarkable precision.

This innovative approach not only enhances the accuracy of scene reconstruction but also introduces novel capabilities for geometric and appearance-based editing. Extensive experiments demonstrate that \method consistently outperforms existing methods in both quality and editing functionality across diverse datasets. These results mark a significant advancement in the domain of interactive 3D editing, setting the stage for future innovations and applications in the field.

{
    \small
    \bibliographystyle{ieeenat_fullname}
    \bibliography{main}
}

\end{document}


\twocolumn[{%
\maketitle
\begin{figure}[H]
\hsize=\textwidth 
\centering
\includegraphics[width=2\linewidth]{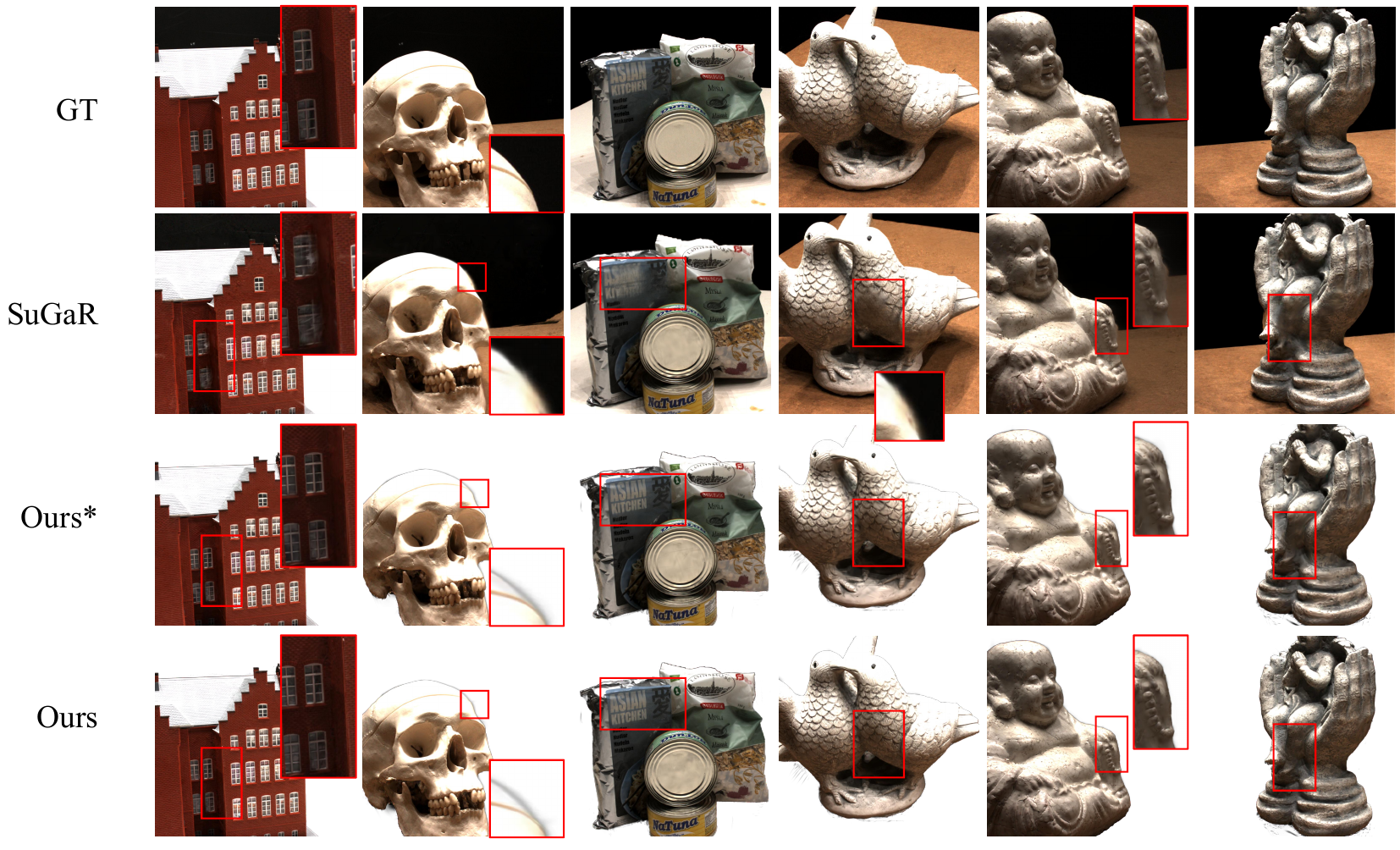}
\caption{\textbf{Qualitative Comparison of View Synthesis on DTU Dataset.}
    Note SuGaR uses a different mesh extraction strategy from ours, so it includes some background.}
\end{figure}
}]

\begin{table*}[tb]
\centering
\resizebox{\linewidth}{!}{
\tabcolsep 3pt
  \begin{tabular}{@{}lcccccccccc@{}}
    \toprule
    \multicolumn{1}{c}{} & \multicolumn{3}{c}{Sugar (1332K)} & \multicolumn{3}{c}{Ours* (39K)} & \multicolumn{3}{c}{Ours (117K)} \\
    \cmidrule(r){2-4} \cmidrule(r){5-7} \cmidrule(r){8-10}
    & PSNR $\uparrow$ & SSIM $\uparrow$ & LPIPS $\downarrow$ & PSNR $\uparrow$ & SSIM $\uparrow$ & LPIPS $\downarrow$ & PSNR $\uparrow$ & SSIM $\uparrow$ & LPIPS $\downarrow$ \\
    \midrule
    24 & 24.02 & 0.871 & 0.177 & 27.83 & 0.895 & 0.158 & \textbf{28.17} & \textbf{0.909} & \textbf{0.112} \\
    37 & 23.36 & 0.849 & 0.225 & 24.67 & 0.862 & 0.163 & \textbf{24.98} & \textbf{0.884} & \textbf{0.128} \\
    40 & 22.69 & 0.824 & 0.284 & 26.78 & 0.843 & 0.250 & \textbf{27.45} & \textbf{0.885} & \textbf{0.177} \\
    55 & 26.97 & 0.794 & 0.311 & 29.46 & 0.936 & 0.106 & \textbf{30.78} & \textbf{0.951} & \textbf{0.073} \\
    63 & 28.61 & 0.921 & 0.185 & 30.61 & 0.949 & 0.105 & \textbf{30.75} & \textbf{0.955} & \textbf{0.084} \\
    65 & 27.67 & 0.805 & 0.328 & 30.83 & \textbf{0.956} & 0.094 & \textbf{30.92} & \textbf{0.956} & \textbf{0.086} \\
    69 & 25.64 & 0.810 & 0.320 & 27.33 & 0.912 & 0.209 & \textbf{27.58} & \textbf{0.921} & \textbf{0.180} \\
    83 & 27.19 & 0.839 & 0.353 & 32.93 & 0.964 & 0.089 & \textbf{33.05} & \textbf{0.967} & \textbf{0.075} \\
    97 & 25.27 & 0.822 & 0.336 & 28.57 & 0.927 & 0.128 & \textbf{28.73} & \textbf{0.928} & \textbf{0.118} \\
    105 & 27.51 & 0.845 & 0.315 & 29.75 & 0.917 & 0.176 & \textbf{30.05} & \textbf{0.931} & \textbf{0.141} \\
    106 & 31.73 & 0.871 & 0.312 & 32.43 & 0.917 & 0.178 & \textbf{33.60} & \textbf{0.938} & \textbf{0.137} \\
    110 & 29.61 & 0.863 & 0.321 & 30.77 & 0.931 & 0.161 & \textbf{31.36} & \textbf{0.938} & \textbf{0.138} \\
    114 & 29.04 & 0.860 & 0.303 & 28.99 & 0.910 & 0.183 & \textbf{29.62} & \textbf{0.925} & \textbf{0.145} \\
    118 & 32.27 & 0.873 & 0.315 & 34.19 & 0.943 & 0.147 & \textbf{35.30} & \textbf{0.957} & \textbf{0.107} \\
    122 & 32.19 & 0.860 & 0.303 & 35.66 & 0.959 & 0.103 & \textbf{36.52} & \textbf{0.966} & \textbf{0.079} \\
    \midrule
    Average & 27.58 & 0.847 & 0.293 & 30.05 & 0.921 & 0.150 & \textbf{30.59} & \textbf{0.934} & \textbf{0.119} \\
    \bottomrule
  \end{tabular}
}
\caption{\textbf{Quantitative Results for View Synthesis on DTU.} In this table, we compare the performance of Sugar and \method on the DTU dataset, with the best results highlighted in bold. Our* denotes the configuration employing our mesh downsampling method with a 1/3 face retention ratio.}
\vspace{-0.5cm}
\label{tab:mesh_dtu}
\end{table*}